\documentclass[smallextended]{svjour3}

\usepackage{graphicx}

\begin{document}

\title{Molecular dynamics study of T=3 capsid assembly}

\author{D.C. Rapaport} 

\institute{Department of Physics, Bar-Ilan University, Ramat-Gan, Israel 52900\\
\email{rapaport@mail.biu.ac.il}}

\date{Received: Oct 2017 / Accepted: }

\maketitle

\begin{abstract}

Molecular dynamics simulation is used to model the self-assembly of polyhedral 
shells containing 180 trapezoidal particles that correspond to the T=3 virus 
capsid. Three kinds of particle, differing only slightly in shape, are used to 
account for the effect of quasi-equivalence. Bond formation between particles is 
reversible and an explicit atomistic solvent is included. Under suitable 
conditions the simulations are able to produce complete shells, with the 
majority of unused particles remaining as monomers, and practically no other 
clusters. There are also no incorrectly assembled clusters. The simulations 
reveal details of intermediate structures along the growth pathway, information 
that is relevant for interpreting experiment.

\end{abstract}

\section{Introduction}

The spontaneous formation of spherical capsids \cite{cri56,cas62} that package 
the genetic payloads of viruses is one of the more fascinating examples of 
supramolecular self-assembly. Capsid shells, in which icosahedral symmetry is 
typically a prominent feature, are constructed from multiple copies of one or a 
small number of different capsomer proteins \cite{bak99}. This symmetry 
simplifies the overall structural organization and minimizes the specifications 
of the construction process, important details since all necessary information 
must be part of the viral genetic payload.

Experimental methods capable of providing sufficient space and time resolution 
for direct observation of the intermediate states constituting the assembly 
pathway have proved elusive until only recently. As a result, knowledge of the 
detailed assembly process is limited, even under controlled {\em in vitro} 
conditions in which capsomers are able to form complete shells without genetic 
material \cite{pre93,cas04,zlo11}. Justification for focusing on this reduced 
version of the process stems from the overall robustness of the self-assembly 
process \cite{cas80}. 

An extensive theoretical corpus covering capsid structure and assembly embraces 
a range of approaches that include thin shells \cite{lid03}, tiling 
\cite{twa04}, particles embedded on spheres \cite{zan04}, elastic networks 
\cite{hic06}, stochastic kinetics \cite{hem06}, nucleation theory \cite{zan06}, 
combinatorics \cite{moi10}, master equations \cite{kee06}, domain decompostion 
\cite{pol13}, and concentration kinetics \cite{mor09,hag10}, the last of which 
is used for interpreting experiment \cite{zlo99,cer02,jon05,van07} and analyzing 
reversible growth \cite{zlo07}. There have been numerous simulations of capsid 
self-assembly using molecular dynamics (MD) \cite{rap99b,rap08b,hag06,ngu07}, 
and Monte Carlo methods \cite{wil07,joh10} that avoid dealing with dynamics; 
modeling capsid assembly is surveyed in \cite{hag14}. Experimental `analog 
simulations' of assembly have been performed using solutions of small plastic 
particles with adhesive-coated surfaces \cite{bre99}.

The work described in this paper is a continuation of a series of MD simulations 
modeling capsid self-assembly with rigid particles. The particles are 
constructed from soft spheres fused together to produce an overall shape 
tailored to form polyhedral shells, given the appropriate interparticle bonding 
interactions. The original simulations \cite{rap99b,rap04a} were severely 
limited by the computational resources available at the time; consequently, the 
focus was on demonstrating the feasibility of assembly in the absence of 
solvent, subject to the condition that the bonding process was irreversible, 
meaning that bonds, once formed, were unbreakable. Shells of size 60 were grown 
from triangular and trapezoidal particles, the latter corresponding to the 
structure of a T=1 virus, as well as shells of size 180 resembling the T=3 
virus. This was followed by a more computationally demanding study based on 
reversible assembly, for T=1 shells only, also described in \cite{rap04a}. In 
the reversible case, bonds are allowed to break when sufficiently stretched. 
Reversibility is clearly more reasonable from a physical perspective, but since 
bonds do not typically undergo spontaneous breakage, the approach required that 
smaller particle clusters be decomposed at regular intervals to avoid kinetic 
traps due to a depleted monomer concentration.

Inclusion of an atomistic solvent became feasible as a result of further 
increases in computer performance \cite{rap08b}. The effect of the solvent is to 
moderate the assembly process, allowing coexisting populations of large clusters 
and monomers, and eliminating the need to artificially decompose small clusters. 
However, computational limits restricted the study to the case of triangular 
particles assembling into 20-particle icosahedral shells. From these simulations 
it became apparent that assembly involves a sequence of reversible steps, with a 
high yield of complete shells and a strong preference for minimum-energy 
intermediates. While ostensibly paradoxical, reversibility is directly 
responsible for effective assembly due to its ability to discourage particle 
clusters from becoming trapped in configurations inconsistent with a successful 
outcome.

Subsequent improvement in computational capabilities led to simulations of 
larger T=1 shells (size 60) in solution \cite{rap12c}. Increased shell size 
allows a broader variety of growth pathways, permitting `entropic' effects to 
compete more effectively with the energetic considerations that select the 
pathways of the smaller shells. Comparing the outcomes of growth simulations 
involving different shell sizes can provide insight into how the size factor 
influences growth and, in particular, which aspects of growth are common to 
different shell sizes. In principle, the observed growth pathways and predicted 
time-dependence of the populations of intermediate structures can be related to 
behavior that could be measured experimentally \cite{end02}, and, as 
experimental technique is refined \cite{med16}, a search for common structural 
features becomes feasible.

The present paper describes self-assembly simulations of T=3 shells from 
trapezoidal particles with reversible bond formation, in the presence of an 
explicit atomistic solvent, and is the natural extension of earlier work on the 
smaller (icosahedral and T=1) shells. The model used for the T=1 case, where all 
trapezoidal particles in the shell occupy equivalent positions, is inadequate 
for T=3 shells where quasi-equivalence \cite{cas62,cas85,ros85} must be 
introduced and, as with the earlier irreversible, solvent-free case 
\cite{rap04a}, a simple method for accommodating this effect is employed. 
Subsequent sections discuss the methodology used for capsomer modeling, 
simulation and analysis, and describe the results.

\section{Methods}

MD simulation of large systems over long time periods requires simplified 
models, avoiding excessive molecular details that would be overwhelming; 
confirmation of a model's adequacy, both qualitative and quantitative, must come 
from outside the simulational framework. Such simplification underlies the 
choice of molecular representation in the present study of T=3 capsid assembly 
with an explicit solvent, where the computational needs are much heavier than 
in the earlier work.

Capsomers are large compact proteins that fit together to form strongly bound 
closed shells. Design of a simplified model particle for use with MD addresses 
two prominent generic characteristics of the capsomer, its overall shape and the 
interparticle bonding forces, while avoiding the complexities associated with 
specific proteins. Here, the particle that represents a capsomer consists of a 
set of soft spheres fused together into a rigid structure approximating the 
shape of a truncated, trapezoidal pyramid (where the interpenetrability of the 
soft spheres is small relative to the particle size). Attractive forces act 
selectively between interaction sites embedded in the lateral walls of the 
particles and are responsible for bonding.

\begin{figure}
\begin{center}
\includegraphics[scale=0.07]{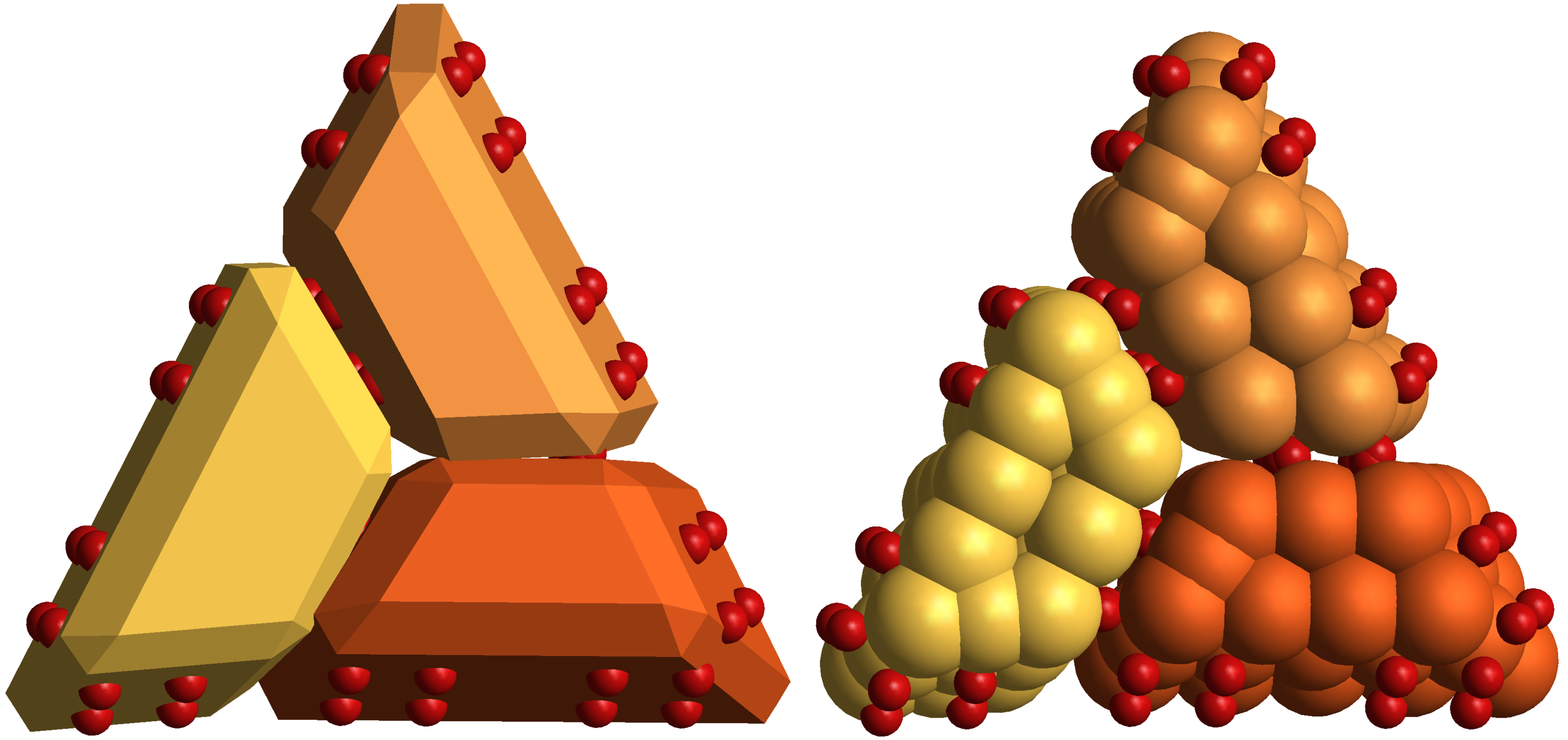}
\end{center}
\caption{\label{fig:fig1}
Bonded trimer showing the effective particle shapes (colors distinguish the 
three particle types) and the bonding sites (in red), as well as the fused 
spheres that form each particle.}
\end{figure}

Unlike the earlier T=1 case, where all particles are identical, T=3 shells are 
are more complicated. Real capsomers are able to undergo small conformational 
changes during assembly, leading to an overall T=3 shell organization referred 
to as quasi-equivalent. In the simulations, for simplicity, the three distinct 
shapes (the differences are slight) are assigned to the particles in advance. 
Figure~\ref{fig:fig1} shows these particles in a fully-bonded trimer 
configuration. The particles have relative dimensions and facet angles 
consistent with the T=3 shell structure. The shell itself is a rhombic 
triacontahedron \cite{wil79} with 30 identical faces (see Figure~\ref{fig:fig2} 
below); each face is subdivided into two isosceles triangles (with base angle
$58.283^\circ$, so the triangles are almost equilateral), and each of these 
triangles is divided into three coplanar trapezoids, yielding a total of 180 
components. The lateral faces of trapezoids in the same triangle, and faces 
between triangles of the rhombus, are normal to the triangle plane, while the 
other lateral faces are inclined at $18^\circ$ (corresponding to a dihedral angle 
of $144^\circ$). These angles are used in specifying sphere coordinates for the 
the three kinds of trapezoidal particles used for shell construction.

The excluded-volume interaction between soft spheres on different particles is 
based on the truncated Lennard-Jones potential
\begin{equation}
u_s(r) = 4 \varepsilon \, [(\sigma / r)^{12} - (\sigma / r)^{6}] + \varepsilon
  \qquad r < r_c \label{eq:ssint}
\end{equation}
where $r$ is the distance between spheres; a small cutoff range, $r_c = 2^{1/6} 
\sigma$, ensures that the force is repulsive. This potential is sufficiently 
stiff to ensure that particle overlap is minimal.

The bonding forces needed for assembly involve quartets of interaction sites on 
the lateral faces of the particles, shown in Figure~\ref{fig:fig1}, and 
act between corresponding sites on specific face pairs of different particles.
The difference between the earlier T=1 case and T=3 is that particle type must 
now be taken into consideration. The use of several interaction sites per face 
ensures that particles are correctly positioned and oriented when in a low energy 
state. The interaction has the form of an inverse power law, gradually changing 
to a stretched harmonic spring below a small crossover distance,
\begin{equation}
u_a(r) = \left\{
\begin{array}{ll}
e (1 / r_a^2 + r^2 / r_h^4 - 2 / r_h^2) & \quad r < r_h \\
e (1 / r_a^2 - 1 / r^2) & \quad r_h \le r < r_a
\end{array}
\right. \label{eq:attrint}
\end{equation}
The overall attraction strength is governed by an adjustable parameter $e$. The 
range of this force, $r_a = 3 \sigma$, is similar to the particle size; the 
crossover distance is $r_h = 0.3 \sigma$, resulting in a narrow harmonic well 
that limits structural fluctuations when in a low energy state. While the 
attraction between sites is not in itself directionally dependent, the 
involvement of several site pairs enforces particle alignment, further enhancing 
the rigidity of multiply-bonded structures. A side-effect of the limited 
structural fluctuations is that the final stages of shell assembly are 
prolonged, because incoming particles must be correctly aligned for entry into 
shell openings that leave minimal space for maneuvering.

The particles themselves are immersed in a neutral solvent formed from the same 
soft-sphere atoms used for the capsomers, a (computationally necessary) 
simplification replacing the (complex) aqueous medium of reality. A thermostat 
is included in the solvent dynamics to control temperature \cite{rap04bk}. 
Although an explicit solvent requires significant additional computation, it has 
several advantages over the implicit (stochastic) alternative. While both are 
capable of serving as heat baths for energy exchange when bonds form and break, 
as well as inhibiting the otherwise ballistic particle motion by adding an 
effective diffusive component ensuring conditions closer to equilibrium, only 
the explicit approach allows particles that have assembled into structures to 
offer mutual shielding against disruptive solvent effects, aids cluster breakup 
without subassemblies needing to collide directly, and incorporates the 
dynamical correlations of the fluid medium. The choice of solvent representation 
also affects the actual particle dynamics and can influence the outcome of 
self-assembly simulations \cite{spa11}.

Typically, relatively large systems and long runs are required to cover the 
multiple length and time scales intrinsic to the system. A run must follow the 
evolution in its entirety, from the initial monomeric state until the expected 
(or unexpected) shells have had the opportunity to self-assemble. Larger shells 
increase the computation time, as well as requiring more particles to produce 
enough shells. The size ratio of the particles relative to the solvent atoms is 
much smaller than in reality, in order to enhance particle mobility; the 
corresponding mass ratio (here 15) is also reduced. Because of the relatively 
high mobility and particle concentration the assembly timescales are highly 
compressed; this is essential to allow the simulation to span the duration of 
the assembly process, and while this may preclude direct quantitative comparison 
with experiment, qualitative aspects of the process, both structural and 
dynamical, ought to be preserved.

Other aspects of the simulations are covered by general MD methodology 
\cite{rap04bk}, including the organization of the force evaluations in a manner 
that scales linearly with the number of particles, dynamics of rigid bodies, 
stable integration of the equations of motion, boundary conditions and 
initialization. Results are expressed in reduced (dimensionless) MD units; these 
are readily converted to physical units (the actual values do not appear in the 
simulation) for comparison with experiment. The reduced unit of length is 
expressed in terms of $\sigma$ (which for argon is 3.4\,\AA\ -- a typical 
value), and solvent spheres have unit mass. Energy is expressed in terms of 
$\varepsilon$, leading to a reduced time unit corresponding to $2.16 \times 
10^{-12}$\,s (also for argon); the integration time step is 0.005. Setting 
$\varepsilon / k_B = 1$ ($k_B$ is the Boltzmann constant) determines the 
temperature unit; a fixed temperature of 0.67 is maintained by the thermostat. 
Additionally, the MD algorithms have been optimized for GPU use \cite{rap11a}. 
These GPU techniques (that, in many respects, differ significantly from their 
conventional counterparts) have been extended to handle the specialized needs of 
the current work, including rigid bodies, multiple particle species, various 
force types, and have also been updated to utilize new hardware capabilities in 
more recent generations of GPUs not available for the earlier work (the GPU used 
here is the NVIDIA K20, whose 2496 computational cores are fully utilized in the 
simulations).

Numerical snapshots of the particle configurations are recorded at intervals of 
$2 \times 10^4$ time steps during the simulation. Interactive visualization 
capabilities are incorporated into the simulations allowing progress to be 
monitored, and also enabling subsequent replay of the recorded snapshot data. 
The snapshots provide the raw data for offline analysis, primarily the search 
for assembled structures by means of cluster analysis, as well as for producing 
imagery such as that included below.

Quantitative cluster analysis is based on an intuitive geometrical definition of 
bonding, namely, if a pair of trapezoidal particles have all four sets of 
matching attraction sites within a prescribed range, $r_b$, they are considered 
bonded. This is merely a bookkeeping device, since there is nothing special 
about bond formation given that bonding is reversible (a key factor for 
successful shell assembly). Once all the bonds have been assigned cluster 
identification is straightforward \cite{rap04bk}, since each connected set of 
bonded particles defines a cluster. Setting $r_b = 0.5$ ($> r_h$) leads to 
results consistent with direct visualization, namely that structural 
fluctuations cause minimal spurious bond breakage and there is no false bond 
designation. Given that the particle design and parameterization ensures that 
bonded pairs have very limited relative motion, the only permitted cluster in 
which every particle has a full complement of five bonded neighbors is a closed 
shell, in the present case having size 180; mutant clusters (whose size would be 
unbounded) do not develop for the range of $e$ considered here.

\section{Results}

The total number of particles in the system is $N = 262144$ ($64^3$), over twice 
that of \cite{rap12c}. There are $N_p = 8650$ particles, divided among the three 
species, sufficient for up to 48 complete shells of size 180; the solvent 
consists of $N - N_p$ atoms, so that the particle fraction is $p = N_p / N = 
0.033$. The overall number density is set to 0.1; this determines the volume of 
the cubic simulation region, and represents a compromise that ensures the 
solvent influences the motion of the particles without excessively impeding it. 
The range of the interaction strength parameter $e$ considered here lies between 
0.086 and 0.089 where the most interesting results appear. Run length is chosen 
to maximize growth products within a tolerable computation time; a single run 
covering $6 \times 10^8$ time steps requires $\sim 30$ days of GPU computation.

\begin{figure}
\begin{center}
\includegraphics[scale=0.2]{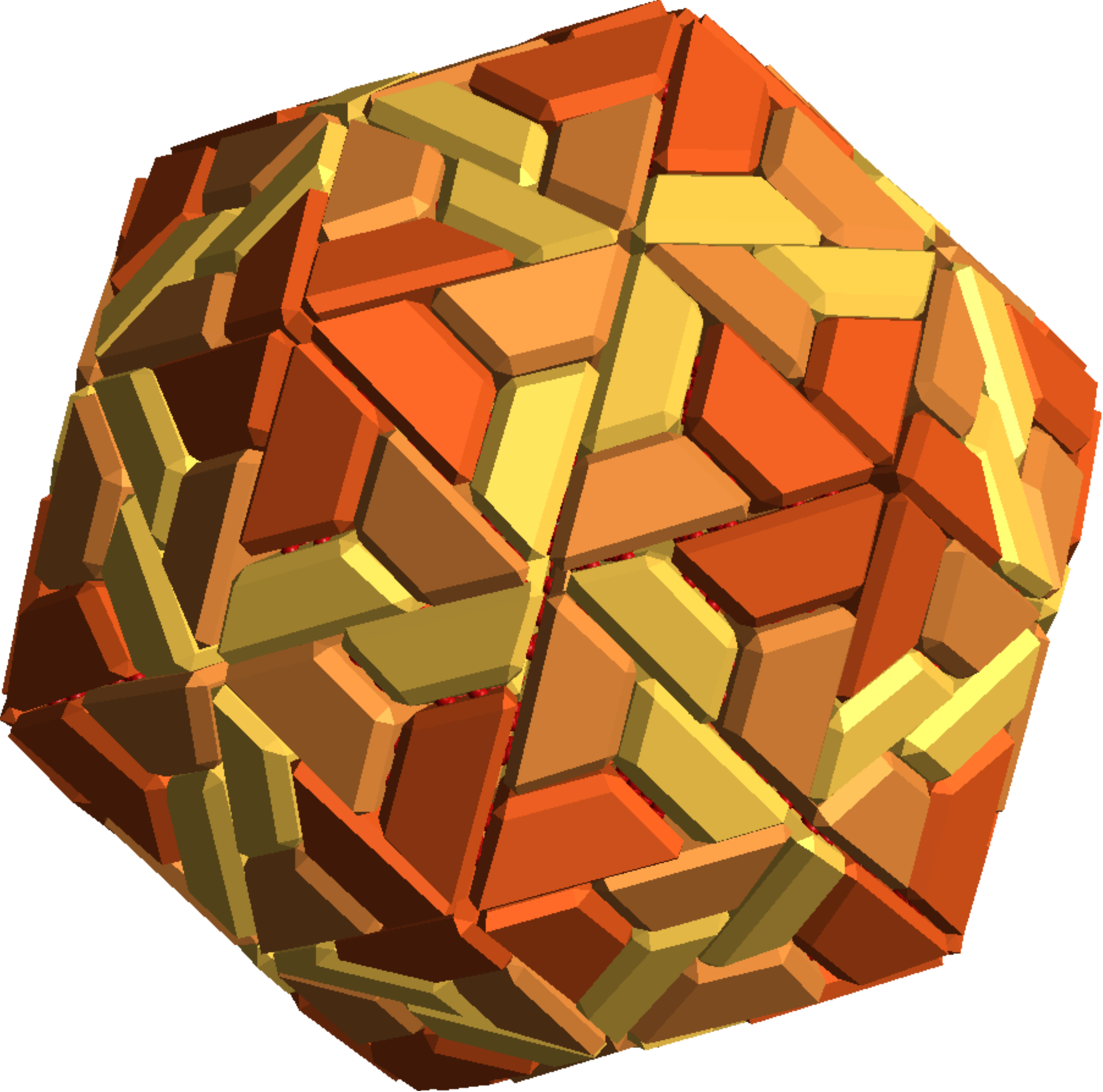}
\end{center}
\caption{\label{fig:fig2}
Single complete shell with 180 particles; colors distinguish the three particle types.}
\end{figure}

Visualization provides the most comprehensive summary, albeit qualitative, of 
the outcome of the simulations. Figure~\ref{fig:fig2} shows a closeup of a 
single complete shell from one of the simulations; the aim is to maximize the 
yield of such shells.

\begin{figure}
\begin{center}
\includegraphics[scale=0.2]{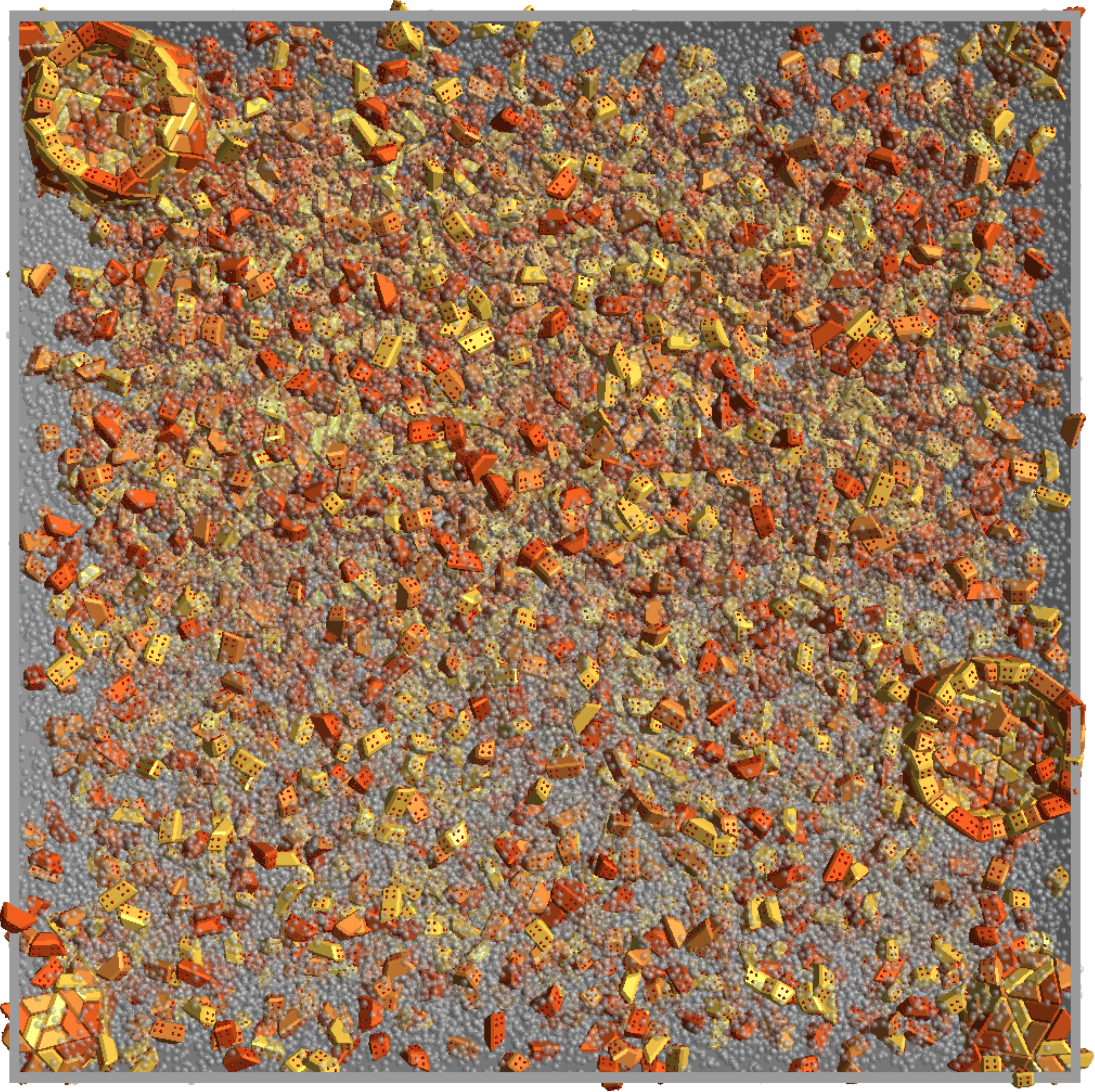}
\end{center}
\caption{\label{fig:fig3}
A picture of the $e = 0.086$ system showing clusters, monomers and solvent near 
the end of the run, with the space-filling solvent shown semitransparently; 
because of periodic boundaries some complete shells appear fragmented at 
opposite faces of the region.}
\end{figure}

\begin{figure}
\begin{center}
\includegraphics[scale=0.2]{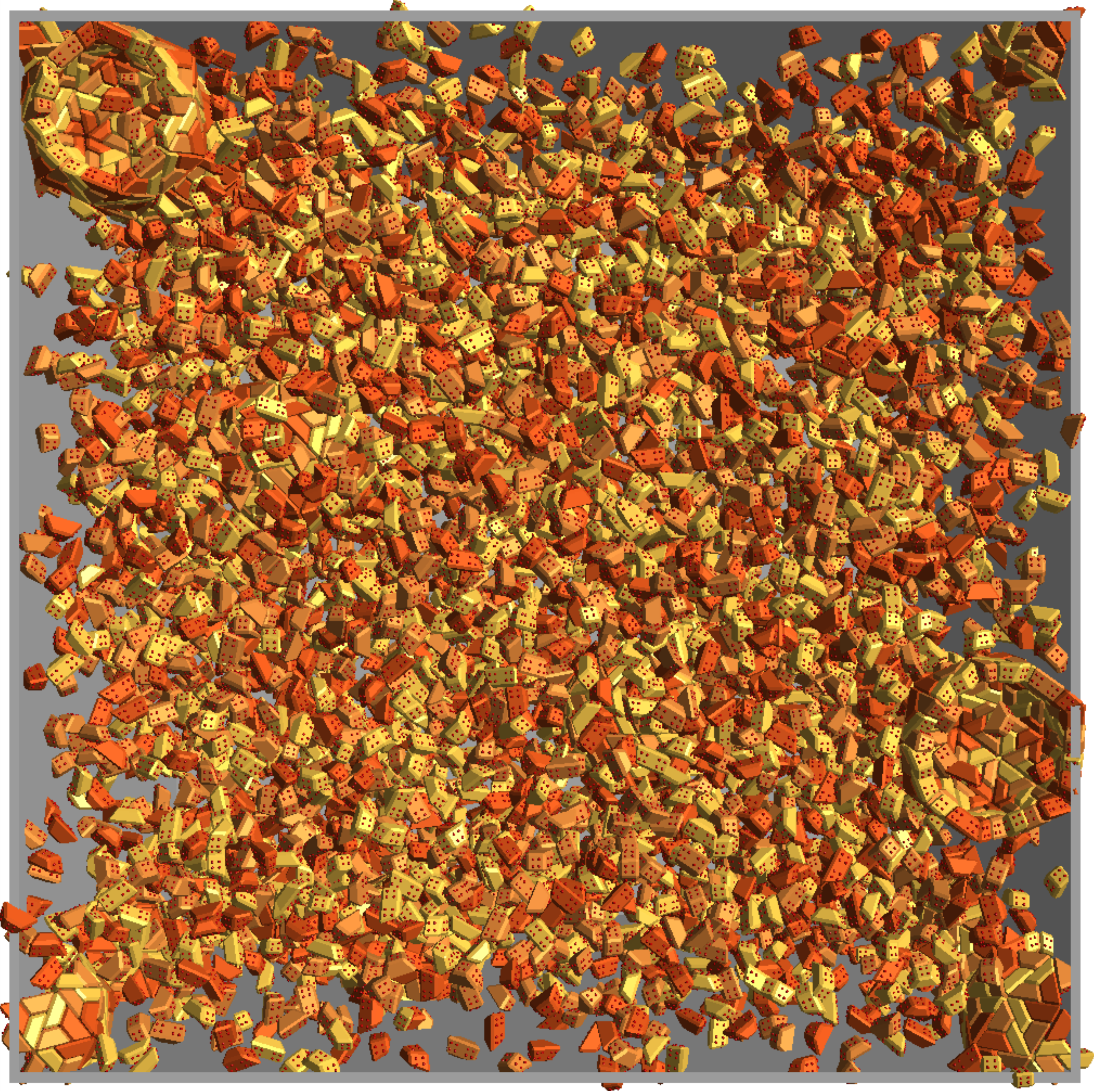}
\end{center}
\caption{\label{fig:fig4}
The same $e = 0.086$ system without the solvent.}
\end{figure}

\begin{figure}
\begin{center}
\includegraphics[scale=0.2]{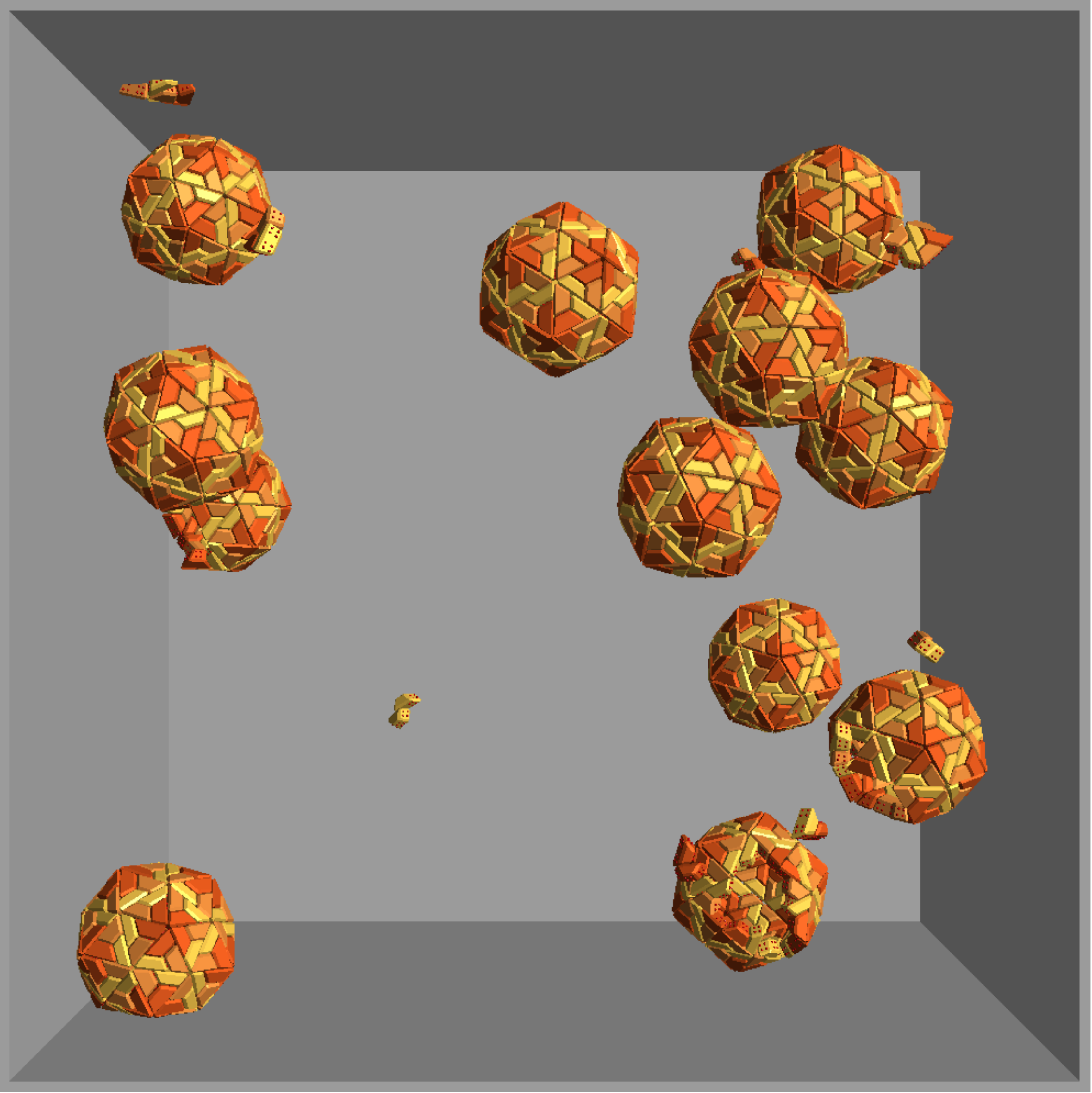}
\end{center}
\caption{\label{fig:fig5}
Another view of the $e = 0.086$ system; both solvent and monomers are omitted 
allowing the clusters to be seen clearly.} 
\end{figure}

Figure~\ref{fig:fig3} shows the complete $e = 0.086$ system close to the end of 
the run, including clusters, monomers and the space-filling solvent; the solvent 
atoms are shown semitransparently to avoid obscuring the interior entirely. The 
same configuration without the solvent appears in Figure~\ref{fig:fig4}. Note 
that clusters that cross periodic boundaries appear as two or more pieces, a 
visualization artifact that can be reduced but not completely avoided by 
translating the system. Another snapshot without the more than 6500 residual 
monomers (which also hide the interior) is shown in Figure~\ref{fig:fig5}; here, 
seven complete shells can be seen, two clusters with 172 and 178 particles, 
another three with size $> 100$, and a few very small clusters.

\begin{figure}
\begin{center}
\includegraphics[scale=0.2]{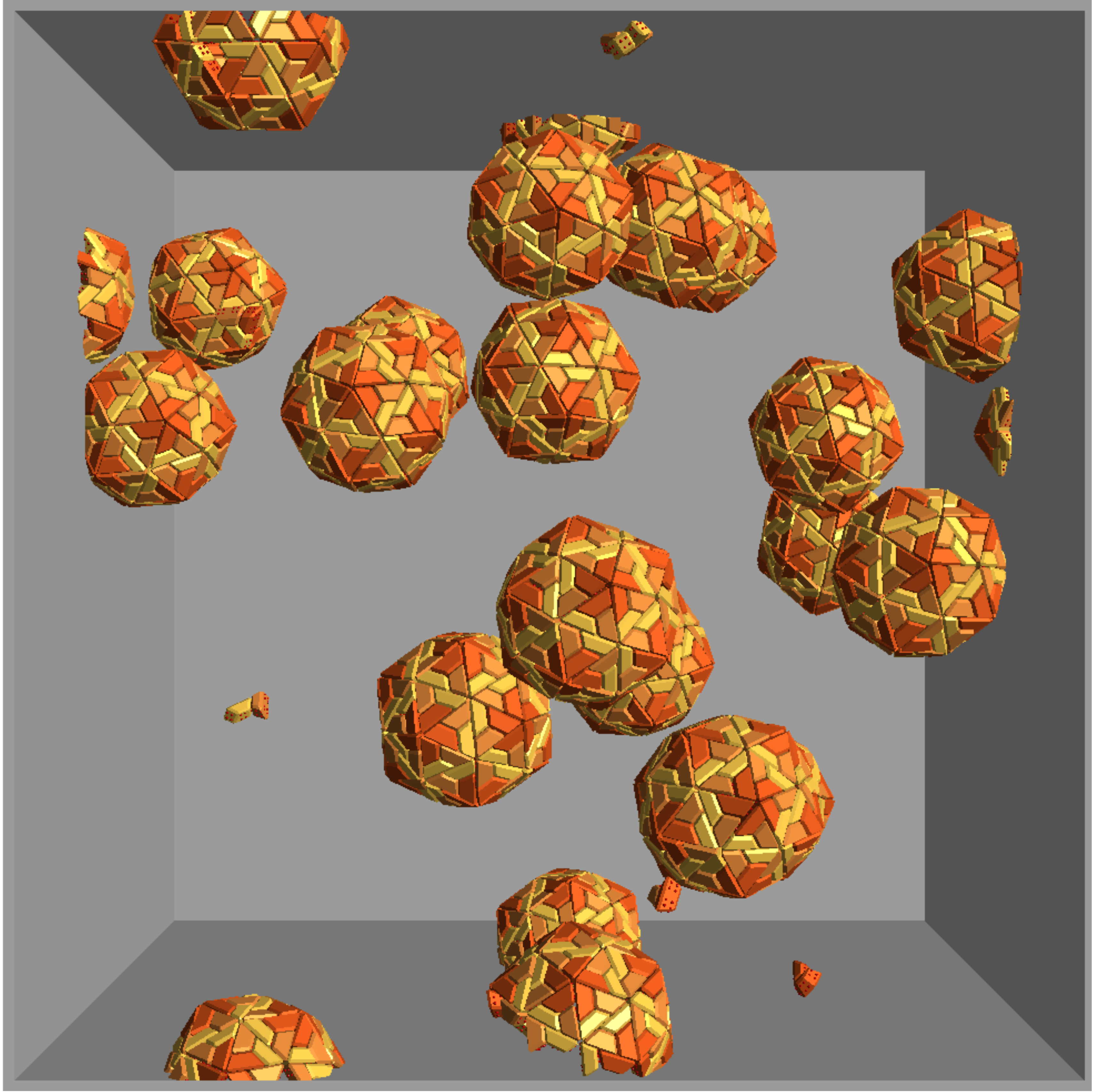}
\end{center}
\caption{\label{fig:fig6}
Clusters at the end of the $e = 0.087$ run.}
\end{figure}

Figure~\ref{fig:fig6} shows the cluster configuration at the end of the $e = 
0.087$ run. Here there are 16 complete shells (out of a possible 48), two of 
size 179, one each of sizes 178 and 118, and four of size $< 7$; there are also 
5100 monomers.

Cluster analysis, based on the approach described above, complements images of 
this kind with the quantitative details of shell production, including how the 
shell yields depend on parameters such as $e$, and the time-dependent aspects of 
assembly. The outcomes of the runs considered here are summarized in 
Table~\ref{tab:mass_frac} where the final cluster mass fractions, grouped 
according to size range, are listed. The tabulated results include the monomer 
fractions, mass fractions of the smallest clusters (range 2--5), intermediate 
size clusters (two ranges, 6--100 and 101--170), almost complete and complete 
clusters ($> 170$), and the complete shells on their own. Subsequent graphs 
(below) provide a detailed visual breakdown of this information, without 
recourse to data grouping. The residual monomer fraction is seen to fall with 
$e$, whereas the fraction of almost complete and complete clusters increases. 
Since almost complete clusters (e.g., size $> 170$) are likely to reach 
completion eventually, this quantity is probably a more useful measure of 
$e$-dependence than the fraction of complete clusters alone, even if the visual 
impact is reduced. Note also the intermediate size (101--170) clusters that 
start to appear at higher $e$ coinciding with falling shell production.

\begin{table}
\caption{\label{tab:mass_frac}
Final cluster mass fractions for different interaction strengths $e$, grouped by 
cluster size or size range.}
\begin{tabular}{lr|cccccc}
\hline
$e$ & Time steps & \multicolumn{6}{c}{Cluster mass fraction} \\
    & ($\times 10^6$) & Size: 1 & 2--5  & 6--100 & 101--170 & 171--180 & 180 \\
\hline
0.086 &   600 &  0.7595 & 0.0018 & 0.0015 & 0.0710 & 0.1661 & 0.1457 \\
0.087 &   600 &  0.5901 & 0.0015 & 0.0000 & 0.0135 & 0.3949 & 0.3329 \\
0.088 &   600 &  0.2768 & 0.0002 & 0.0088 & 0.1557 & 0.5585 & 0.3121 \\
0.089 &   570 &  0.2243 & 0.0002 & 0.0055 & 0.1733 & 0.5966 & 0.2289 \\
\hline
\end{tabular}
\end{table}

\begin{figure}
\begin{center}
\includegraphics[scale=0.9]{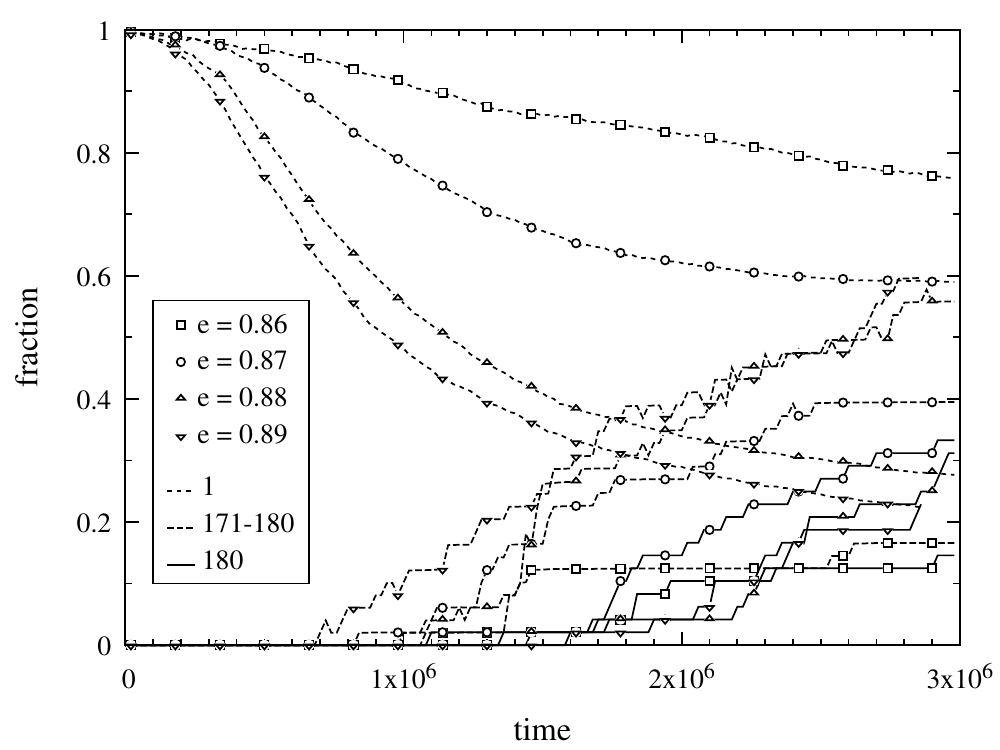}
\end{center}
\caption{\label{fig:fig7}
Time-dependent mass fractions of complete shells (180), complete and almost 
complete shells (171-180) and monomers (1), for different $e$.}
\end{figure}

Figure~\ref{fig:fig7} shows how three of the quantities associated with cluster 
growth vary with time for the different $e$. The first set of curves shows the 
gradual appearance of complete shells after an initial lag time. This is 
accompanied by the diminishing monomer populations shown in the second set of 
curves. Because the final growth steps tend to occur very slowly, the third set 
of curves shows the mass fractions of clusters in the size range 171--180; these 
values, which include the complete shells, are less sensitive to fluctuations 
over different runs (with the same parameters but with different initial states) 
and, as suggested previously, are useful for determining the dependence on $e$ 
since clusters in this size range are likely to form closed shells eventually.

Each shell has a distinct growth pathway through a multitude of possible 
intermediate subassemblies. Automated tracking of individual clusters is 
complicated by their mobility in solution and the fact that particles join and 
leave the cluster in the course of the growth process. Cluster identity is 
therefore based on comparison with a reference state consisting of the set of 
structures in the final configuration, and the cluster with the most member 
particles in common with the final shell is identified as the precursor of that 
shell and given its identity. Since growth generally occurs by addition of 
single particles rather than merging of substantial subassemblies (not 
impossible, but rarely observed), once a subassembly is large enough to achieve 
longevity, cluster identification normally yields an unambiguous result 
consistent with visual monitoring. Problems can occur with very small clusters, 
but these are of minor interest when considering the overall growth process; 
examining these small clusters is a separate issue when considering how growth 
is initiated.

\begin{figure}
\begin{center}
\includegraphics[scale=0.9]{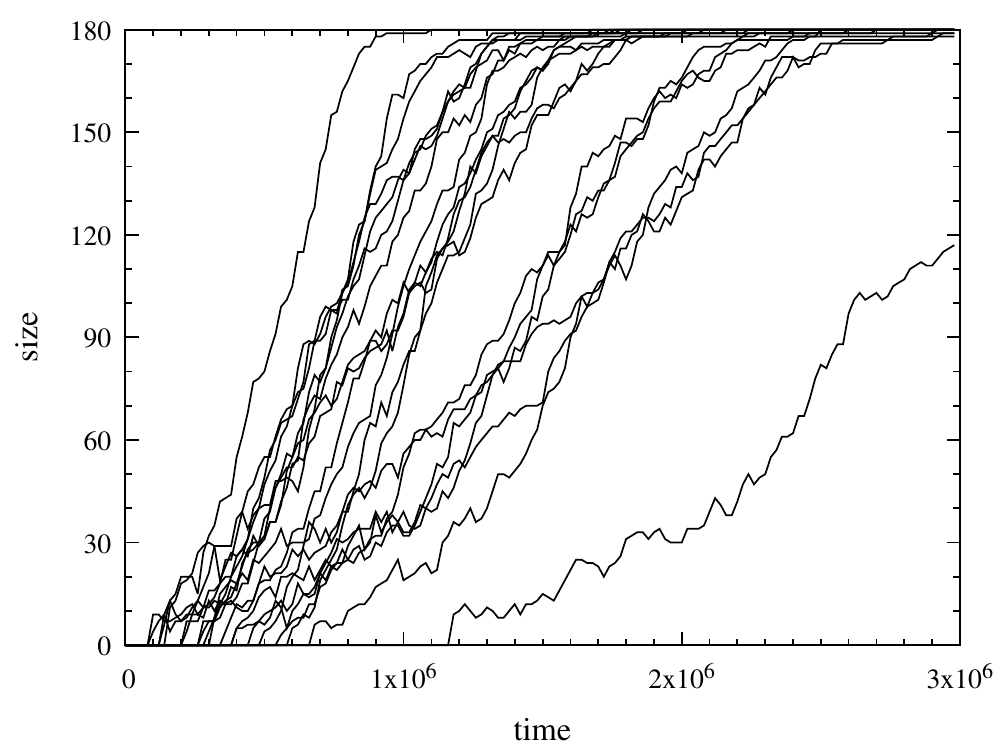}
\end{center}
\caption{\label{fig:fig8}
Graph showing growth of individual clusters, for $e = 0.087$.}
\end{figure}

Figure~\ref{fig:fig8} graphs the development of individual clusters for $e = 
0.087$, almost all of which grow into complete shells. The different curves show 
a considerable spread in growth rates and the lack of monotonicity in the curves 
is evidence that self-assembly pathways are not unidirectional; the spacing of 
the configurational snapshots determines the temporal resolution. Other particles 
not in the final shell will also join the cluster temporarily, typically only 
for short periods, but are excluded from this analysis.

\begin{figure}
\begin{center}
\begin{tabular}{c}
\includegraphics[scale=0.6]{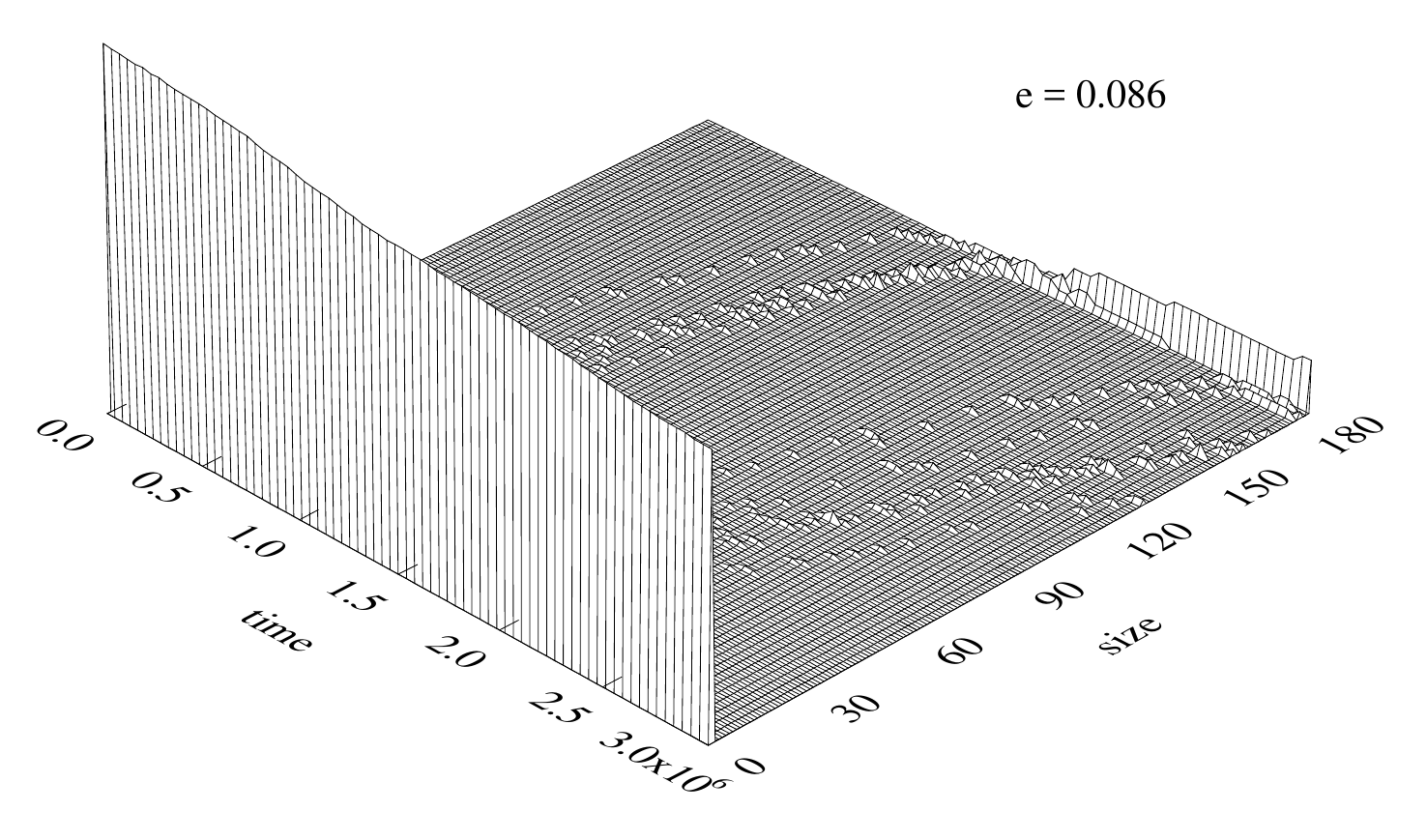} \\[-15pt]
\includegraphics[scale=0.6]{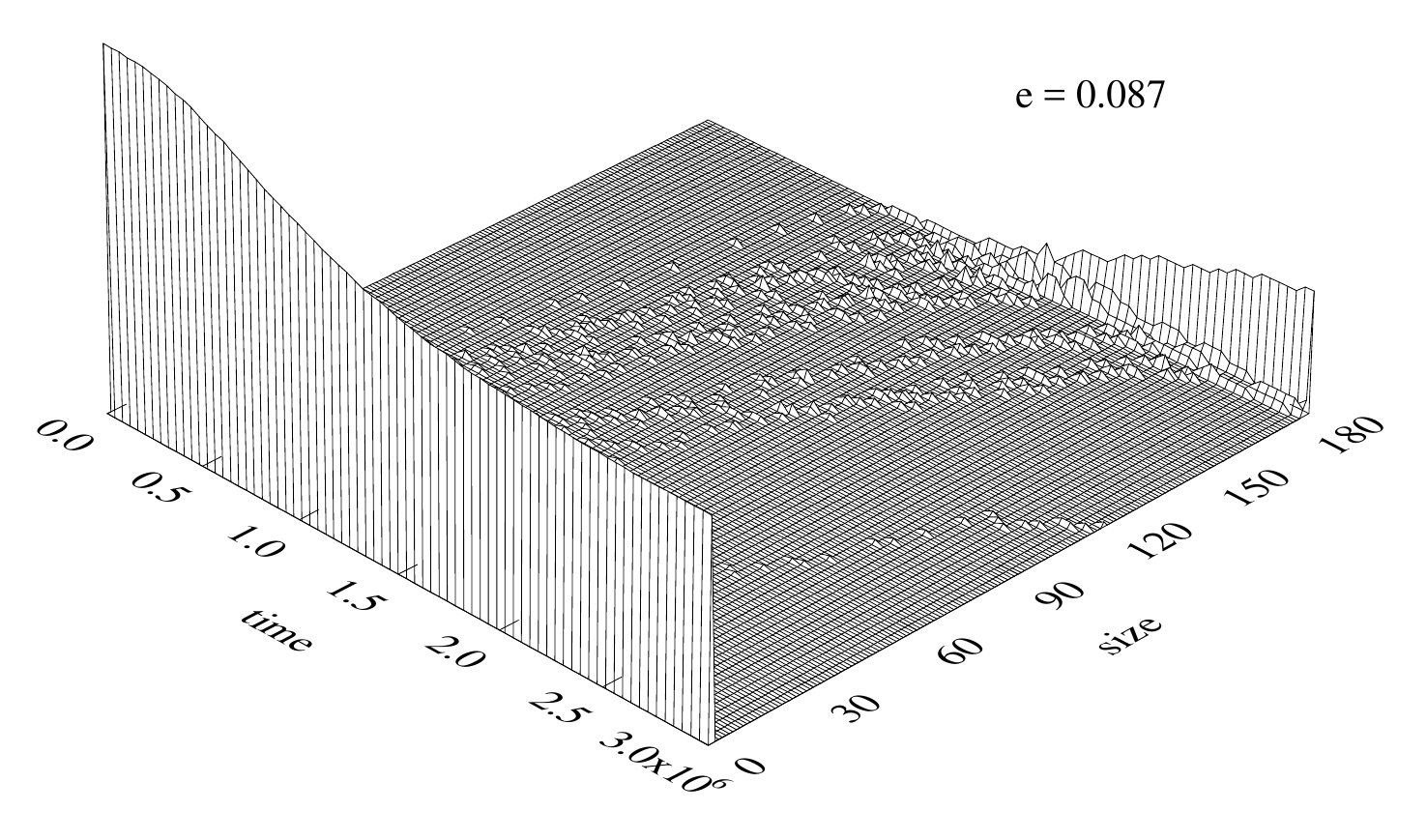} \\[-15pt]
\includegraphics[scale=0.6]{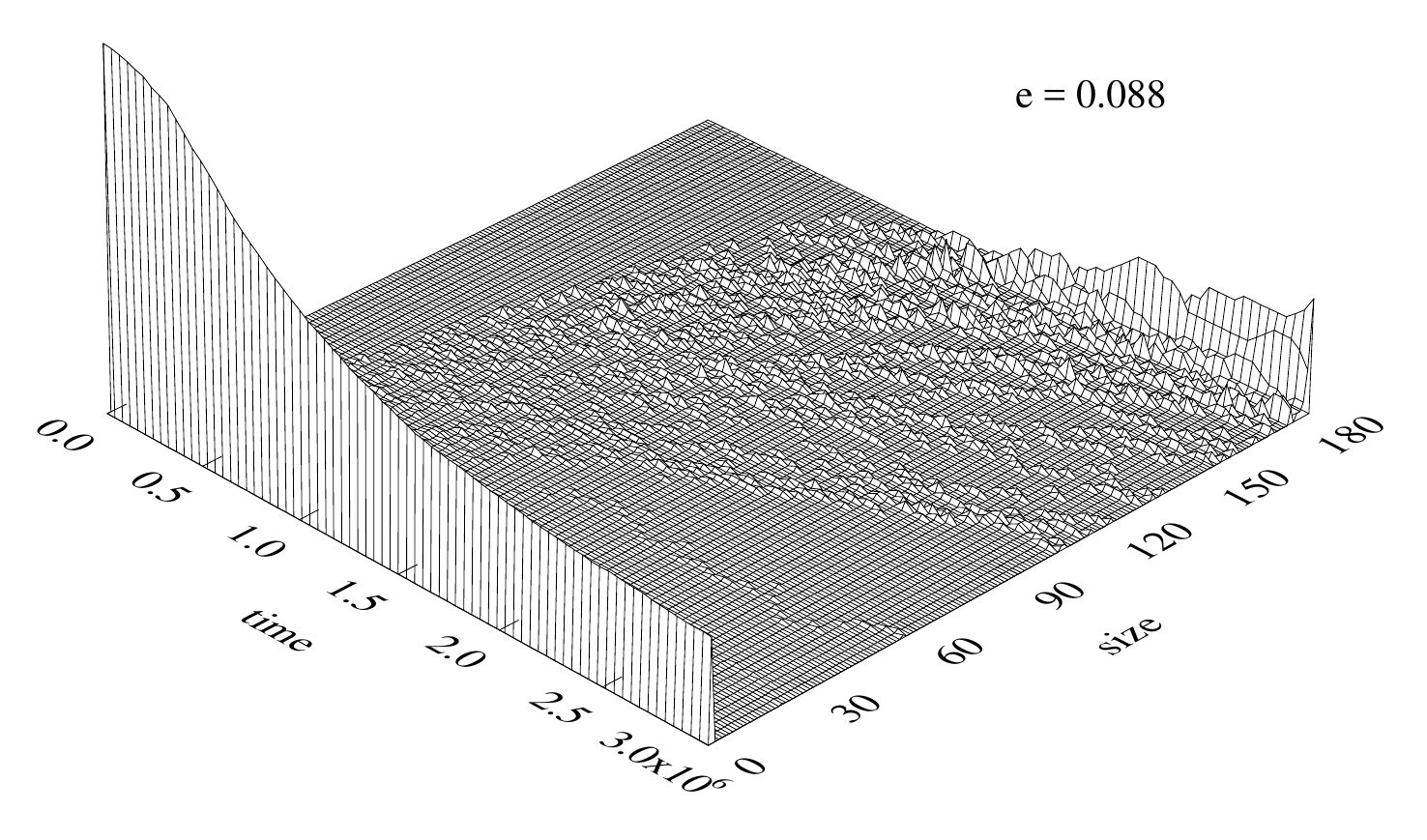} \\[-15pt]
\includegraphics[scale=0.6]{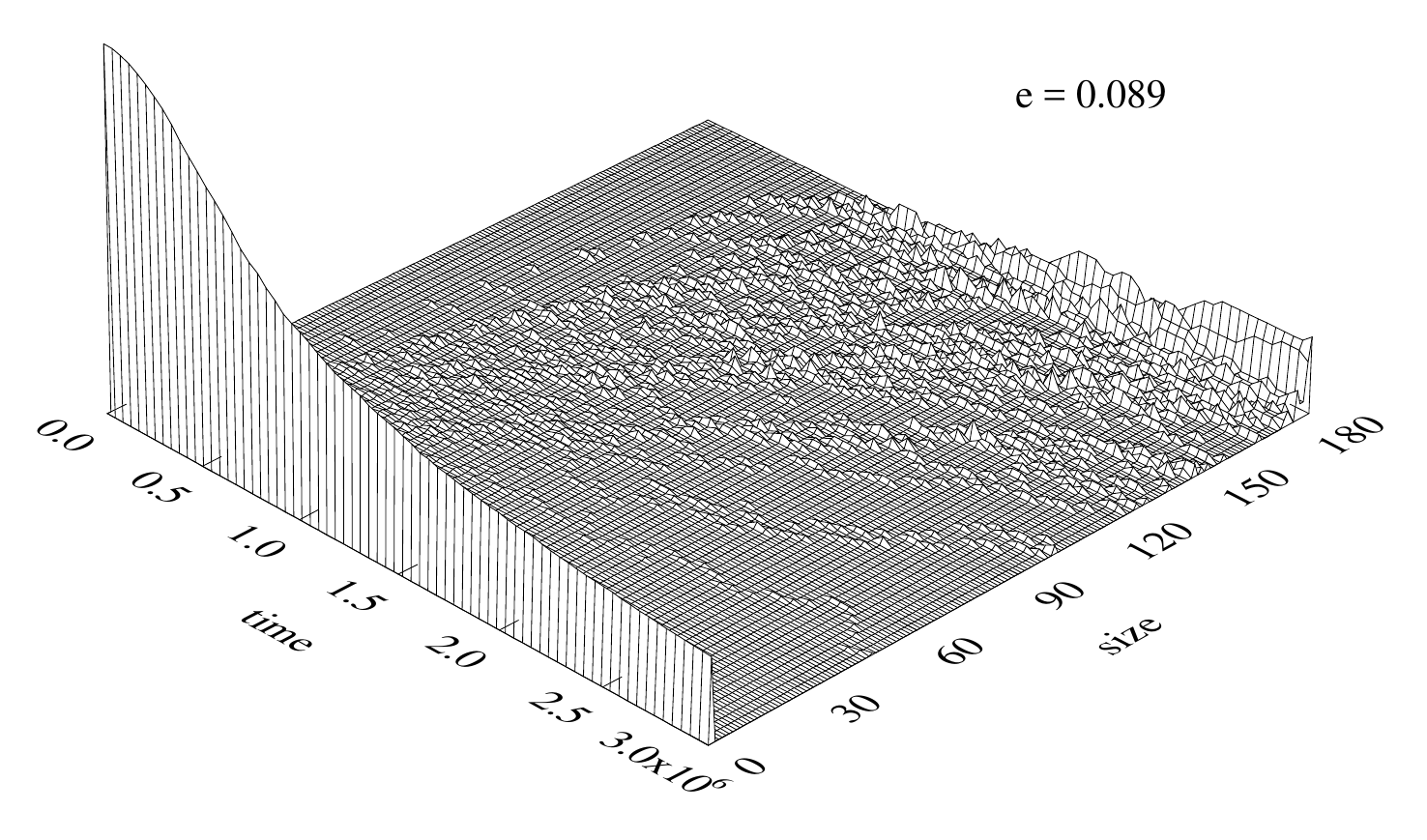} \\
\end{tabular}
\end{center}
\caption{\label{fig:grow_surf_plots}
Time-dependent cluster mass distributions for different $e$.}
\end{figure}

The most detailed quantitative description of how overall cluster growth 
progresses can be summarized using 3D surface plots showing the time-dependent 
mass distributions over the full range of sizes, without any grouping. 
Figure~\ref{fig:grow_surf_plots} shows such plots for each of the $e$ values 
considered here. Comparing the graphs reveals the behavioral trend as $e$ is 
increased, namely a gradual population shift in the final state from monomers to 
shells, and the absence of intermediate size clusters as the process nears 
completion. During the period over which most of the growth occurs the size 
distribution is relatively broad and ill-defined, a consequence of the 
considerable variation in the growth histories of individual clusters. Signs of 
late-developing clusters can be seen.

For $e$ values below the range discussed here (not shown) the population 
distribution shifts towards monomers since the assembly process cannot even get 
started (the stability and lifetimes of dimers and other small clusters were 
studied for T=1 in \cite{rap12c}). At higher $e$ (also not shown) more 
subassemblies are able to grow and the competition for monomers reduces their 
supply below a useful concentration prior to shell completion. Thus there are 
both fewer monomers and an increased population of almost complete shells, while 
the number of complete shells is reduced; even though large clusters can release 
particles back into solution this is usually unhelpful for the growth of other 
shells given the relative slow monomer diffusion rate. Each of these aspects of 
the behavior is a consequence of reversible bonding and all are entirely 
reasonable.

Due to the plethora of intermediate structures there is no unique quantitative 
characterization useful for specifying pathways that could serve as a `reaction 
coordinate'. Several such measures could be introduced, with varying degrees of 
meaningfulness, for different stages of assembly, such as the number of holes, 
hole shapes and sizes, the total length of the boundary surrounding the (one or 
more) holes, and boundary raggedness. Some are more easily formulated and 
evaluated than others, but rather than describe such an exercise, a few images 
of partial shells will be used to illustrate the kinds of intermediate 
configurations observed. The imagery is augmented by the use of color-coding to 
convey details of the attachment sequence; this provides a concise visual 
summary of shell growth history over an extended time interval.

\begin{figure}
\begin{center}
\includegraphics[scale=0.1]{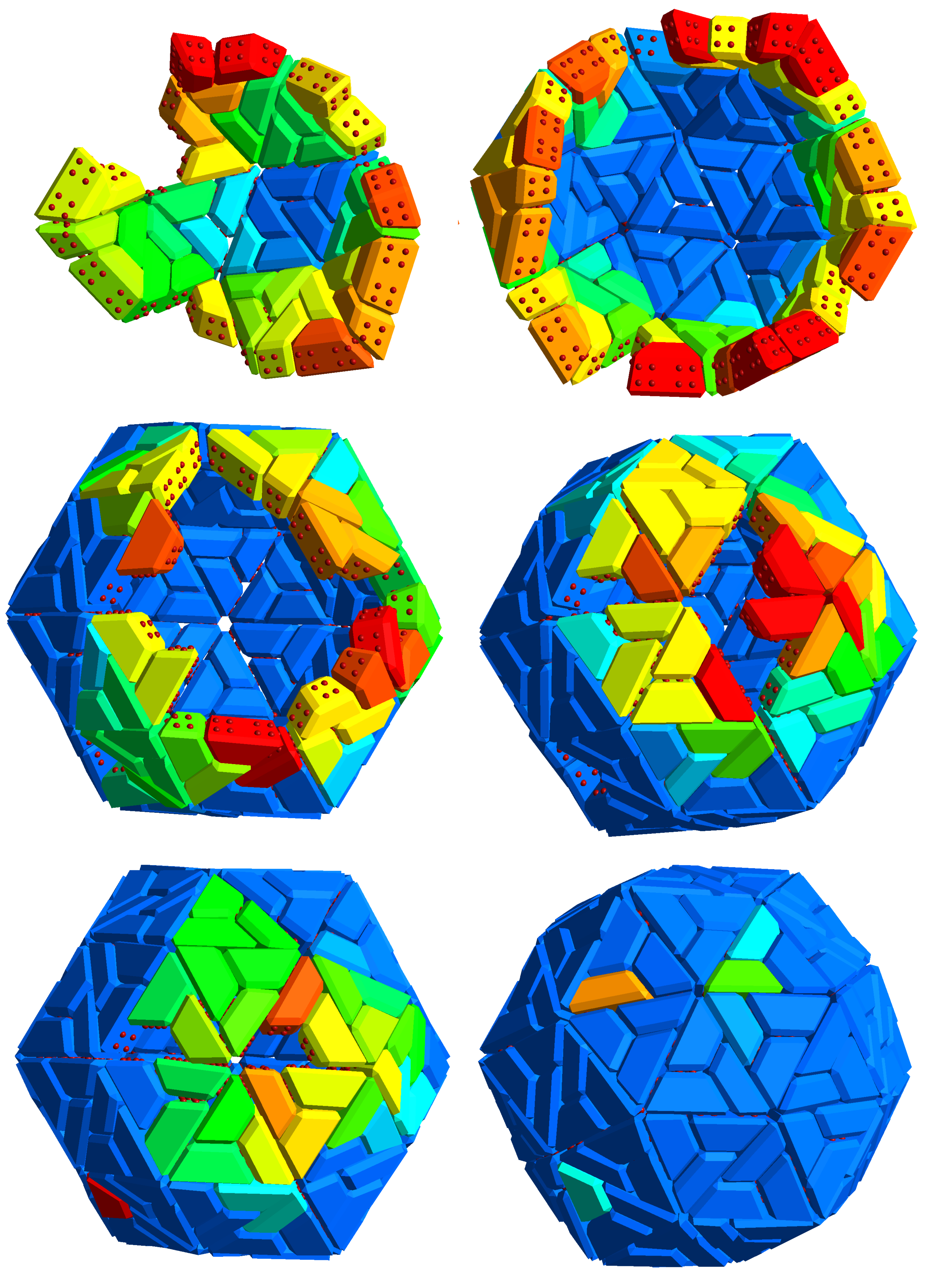}
\end{center}
\caption{\label{fig:fig10}
Color-coded pictures of a single growing cluster; color indicates when particles 
joined the cluster (relative to the current time, ranging from red for most 
recent, to blue for earliest).}
\end{figure}

Individual clusters are followed, using the same automated tracking scheme as 
before, and particles are colored according to when they joined the cluster (or 
most recently joined if there were multiple events of this kind); the colors 
span a sliding range covering the previous $5 \times 10^7$ time steps in the 
cluster's history, starting with red for the most recent additions, through a 
series of spectral colors, to blue corresponding to additions that occurred 
either near the start of the range or at an even earlier time. A series of 
images of this kind, for just a single cluster ($e = 0.087$), appears in 
Figure~\ref{fig:fig10}.

The pictures include examples of the kinds of structural features listed above. 
There is typically just a single major opening, although several small holes can 
remain in later growth stages that fill eventually. The nature of the boundary 
surrounding the particles of the cluster varies in its degree of raggedness. 
Note that only particles that are present in the final shell are shown; 
temporarily bonded particles are omitted, although those that return and 
eventually join the shell are included, with the color coding based on the most 
recent bonding event. Since each cluster has a different history (best seen when 
watched as a movie), the rich variety of behavior is evidence of the difficulty 
in attempting to quantify the finer characteristics of shell growth. Lack of 
quantification should not, however, prevent the comparison of images of this 
type with experiment.

It is interesting to note that similar behavior was originally seen for 
icosahedral shells \cite{rap08b}, almost an order of magnitude smaller, as well 
as for T=1 shells \cite{rap12c,rap14b}. The population distribution is 
essentially binary, either complete shells or monomers, and even small clusters 
that might appear to have enhanced stability, particularly the triangle in 
Figure~\ref{fig:fig1}, are absent under conditions favoring optimal growth. There 
is also the preference for compact structures during early stages of growth 
followed by clusters with boundaries that are increasingly ragged for larger 
shell sizes; in the icosahedral case, where a complete classification of all 
possible intermediates was possible, the clusters were notable for the dominance 
of (near-)maximally bonded states over a much larger number of possible 
alternatives.

In the same way that individual shells can grow in different ways, the mean 
behavior over independent runs can vary. The variation will be prominent during 
the interval when most of the assembly occurs, as well as in the final shell 
yields. These are due to two rate-limiting growth stages: First there is the low 
probability of successfully advancing cluster growth beyond the smallest 
clusters (e.g., dimers and trimers); the fact that small clusters have a low 
survival probability means that successfully initiating the growth of an 
individual shell is a rare event whose consequences affect the entire growth 
process. Second, the extended duration of the final steps to shell completion 
due to the difficulty of inserting the final monomers into one or more small 
holes remaining in the shell. Growth is also retarded because of the serious 
depletion of the monomer population during assembly. Two T=1 runs for one system 
size and the corresponding run for a system double the size were compared in 
\cite{rap14b}. The two smaller systems showed different time-dependent growth, 
while that of the larger lay in between the two. This demonstrates that the final 
yield is sensitive to statistical noise and multiple runs would be needed for 
quantitative studies; these results also suggest that the system size is adequate.

Several other aspects of these self-assembly simulations that were examined in 
earlier work will be mentioned for completeness; corresponding studies have yet 
to be carried for T=3 (or even T=1) with solvent, although the observations are 
likely to remain relevant for the present case. The effect of introducing 
different particle interface energies to encourage formation of intermediates 
(typically dimers or trimers) as a precursor to shell growth and its effect on 
yield and intermediate structures was studied in \cite{rap04a}, for T=1 shells 
without solvent. Closed shells were found to show enhanced stability and breakup 
was not observed, a consequence of particles being restrained by bonding through 
all their lateral faces, and the absence of structural fluctuations capable of 
breaking individual bonds to initiate structural failure. This hysteresis was 
demonstrated for icosahedral shells \cite{rap08b} by showing that if $e$ was 
reduced during the run to a level at which assembly would not have occurred, all 
incomplete assemblies quickly disassembled, leaving only the closed shells.

The appearance of complete shells is confined to a relatively narrow range of 
attraction strengths $e$ (where raising $e$ over a narrow range is equivalent to 
reducing temperature); the range is actually so narrow that the high-yield 
phenomenon could easily have gone unnoticed, sandwiched as it is between the 
region of no growth and that of many large incomplete clusters. The relevant 
range depends on the particle concentration $p = N_p / N$ \cite{rap12c}, a 
parameter that governs the frequency of monomer encounters (that oppose 
solvent-induced cluster breakup); studying this behavior would lead to an 
$e$--$p$ phase diagram, a computationally intensive task given the long runs and 
the need for multiple runs to reduce shell-yield fluctuations. Other parameters 
yet to be studied systematically include overall density and relative particle 
mass, as well as design features of the models (such as the choice of 
interactions, positioning of bonding sites, etc.).

\section{Conclusion}

Earlier simulational studies of shell growth aimed at modeling {\sl in vitro} 
viral capsid growth have been extended to the case of 180-particle T=3 shells. 
Three slightly different particle geometries are used to mimic the effects of 
quasi-equivalence required for shells with more than 60 elements. In agreement 
with previous observations, complete shells can be grown if the parameters are 
correctly chosen, although the computation times required to cover the prolonged 
growth periods and larger systems are substantially longer.

Maximizing the yield of complete shells was an important goal in formulating the 
model; the fact that the region size is limited makes this a more prominent 
issue than it would be {\em in vivo} where other considerations of a more 
biological nature are involved. Since allowing bond breakage might be expected 
to reduce efficiency, the approach used both in the original simulations 
\cite{rap99b} and as one of the alternatives in \cite{rap04a} (see also 
\cite{rap10b}) was to make bond formation irreversible (accomplished by altering 
the form of the pair attraction once inside a suitably defined bonding range, 
together with a complicated procedure aimed at avoiding bonds incompatible with 
the final structure). In practice, it turns out that not only is reversible 
bonding much simpler from a computational point of view (with incompatible bonds 
managing to break unassisted) but, paradoxically, reversibility is a key 
contributor to efficient assembly \cite{rap08b,rap12c,rap14b}. Indeed, 
reversibility constitutes a major difference between assembly at microscopic and 
macroscopic scales, and is a consequence of the thermal `noise' that competes 
with the forces driving growth, an effect absent at the macroscopic scale.

The author declares that there is no conflict of interest.

\bibliographystyle{spmpsci}

\bibliography{trieste}

\end{document}